\documentclass[aps,prl,twocolumn,showpacs,preprintnumbers,superscriptaddress]{revtex4}

\usepackage{graphicx}
\usepackage{epsf} 
\usepackage{dcolumn}
\usepackage{bm}
\usepackage{amssymb}
\usepackage{amsmath}

\begin{document}

\title{Rare Events and Scale--Invariant Dynamics of Perturbations in
Delayed Dynamical Systems}

\author{Alejandro D. S\'anchez}
\author{Juan M. L\'opez}
\email{lopez@ifca.unican.es}
\author{Miguel A. Rodr\'{\i}guez}
\affiliation{Instituto de F\'{\i}sica de Cantabria (CSIC-UC),
E-39005 Santander, Spain}
\author{Manuel A.\ Mat\'{\i}as}
\affiliation{Instituto Mediterr\'aneo de Estudios Avanzados, IMEDEA
(CSIC-UIB), E-07122 Palma de Mallorca, Spain}

\date{\today}

\begin{abstract}
We study the dynamics of perturbations in time delayed dynamical
systems. Using a suitable space-time coordinate transformation, we
find that the time evolution of the linearized perturbations
(Lyapunov vector) can be described by the linear Zhang surface
growth model [Y.-C. Zhang, J. Phys. France {\bf 51}, 2129 (1990)],
which is known to describe surface roughening driven by power-law
distributed noise. As a consequence, Lyapunov vector dynamics is
dominated by rare random events that lead to non-Gaussian
fluctuations and multiscaling properties.
\end{abstract}


\maketitle

Deterministic chaos is an ubiquitous feature of nonlinear systems
and much is understood about its dynamical properties in
low-dimensional systems. Lyapunov exponents, measuring the
separation of initially nearby trajectories, have demonstrated to
be helpful in characterizing chaotic motion in systems with a few
degrees of freedom \cite{schuster}. In the case of spatially
extended nonlinear systems, a straightforward generalization leads
to the concept of Lyapunov exponent density or spectrum, but the
problem here is more complex and one must take into account
propagation and diffusion phenomena that complicate much the
problem \cite{bohr}.

Pikovsky {\it et al.} \cite{pik1,pik2} have recently shown that
the Lyapunov vector, which contains the spatial dependence of the
diverging growth rate of the linearized perturbations, can be
described as a nonequilibrium growing interface. Moreover, they
have reported that in many cases, as for instance in coupled-map
lattices, Lyapunov vector growth is consistent with that of the
well-known Kardar-Parisi-Zhang (KPZ) equation \cite{kpz} of
surface growth. In the case of systems with a conservation law,
one has to be more cautious since KPZ scaling might be not
satisfied, e.g., for the Hamiltonian lattices studied in Ref.
\cite{pik3}. An interesting question that naturally arises is
whether the dynamics of perturbations in extended systems could be
divided in a few universality classes of growth according to the
existence of symmetries and/or conservation laws, as occurs in
nonequilibrium surface growth \cite{baraba&stanley}.

In this Letter we study Delayed Dynamical Systems
(DDSs), which are formally infinite dimensional dynamical systems
and show many aspects of space-time chaos. DDSs appear in a number
of biological and physical situations, e.g., regulation of blood
cells production \cite{mg}, lasers with delayed feedback
\cite{ikeda,arecchi1}, etc. In recent years, DDSs have attracted a
renewed interest as they have been proposed as candidates for
secure communication systems based upon semiconductor lasers with
time-delayed optical feedback \cite{roy,goedgebuer}.

Our aim in this Letter is to show that the dynamics of the
Lyapunov vector in DDSs can be generically described by a
stochastic surface growth equation in which rare fluctuations play
a most important role. We argue that the growth model belongs to
the universality class of the linear Zhang model
\cite{zhang1,krug,baraba,lopez,baraba&stanley}. This is in
contrast with other spatially extended dynamical systems that in
general lead to KPZ-like growth of the amplitude perturbations
\cite{pik1,pik2}. Our results clearly show that, although DDSs are
usually interpreted as extended dynamical systems with
hyper-chaotic behavior, they exhibit distinctive features,
including multiscaling and intermittent driving fluctuations, not
shared with other space-time chaos systems.

A broad class of systems with delayed feedback are given by
differential-delay equations such as
\begin{equation}
\label{dds} \dot{y}={{d y}\over{d t}}={\cal F} (y,y_\tau),
\end{equation}
where $y_\tau=y(t-\tau)$ is the delayed variable and $\tau$ is
time delay. A canonical example is the Mackey-Glass model
\cite{mg} (see below), for which it has been shown that the number
of positive Lyapunov exponents grows linearly with the time delay
\cite{farmer}.

\paragraph{Generic surface growth model for DDSs.--}
Any DDS can be interpreted as a spatially extended system by
introducing a simple transformation, $t=x+\theta\,\tau$, where
$x\in [0,\tau]$ is the {\em space variable}, while $\theta\in
\mathbb{N}$ is a discrete time variable \cite{arecchi2}. This is a
powerful representation in which the formation and propagation of
structures, defects and spatiotemporal intermittency can be more
easily identified \cite{arecchi2}. Note that the time delay $\tau$
becomes the {\em system size} in such a way that the time
dependence with the delayed variable is transformed into a
short-range interaction within the horizontal space coordinate $x$
in the space-time representation.

The dynamics of propagation of localized disturbances, {\it i.e.}
the Lyapunov vector, can be obtained by linearizing (\ref{dds})
which, after the transformation to space-time representation,
becomes \cite{politispace}
\begin{equation}
\nabla\phi(x,\theta)=u\,\phi(x,\theta)+v\,\phi(x,\theta-1),
\label{eqdec1}
\end{equation}
where $\nabla = \partial_x$, $\phi(x,\theta) = \delta y(x,\theta)$
is the Lyapunov vector, and $u={\partial \cal F} / {\partial y}$
and $v={\partial \cal F} / {\partial y_{\tau}}$.

Following Pikovsky {\it et. al.} \cite{pik1,pik2}, we can map this
equation into a surface growth model by introducing the surface
height as $h(x,\theta)=\log\vert\phi(x,\theta)\vert$. Then, after
defining $\partial_{\theta} \phi(x,\theta) = \phi(x,\theta+1) -
\phi(x,\theta)$, one easily obtains the corresponding growth
equation for the surface. We finally arrive to the equation,
\begin{equation}
\partial_\theta h =-(1/v)\, \nabla h + (u / v) +
1\label{interfeqn},
\end{equation}
where $\nabla h = \partial_x h$ and the boundary conditions
$h(\tau,\theta) = h(0,\theta+1)$.

In the following we restrict ourselves to the most studied case of
DDSs in the chaotic regime, in which $u = {\rm const}$ and $v$ is
a nonlinear function (see Eq.(\ref{class}) below), so that $1/v$
will play the role of a noise source. As we shall see below,
probability distribution and correlation of this noisy term will
turn out to be essential for understanding the behavior of the
growth equation. Note that, if $P_v(v)$ is the probability density
function (PDF) of $v$, the corresponding PDF of $\zeta = 1/v$ is
given by $P_{\zeta}(\zeta) = \vert \zeta \vert^{-2}P_v(1/\zeta)$.
Therefore, if $P_v$ behaves as $P_v(v) \sim \vert v \vert^{\nu}$
when $v \to 0$, we have that the PDF of $\zeta$ is then given by
$P_{\zeta}(\zeta) \sim \vert \zeta \vert^{-(2+\nu)}$ for large
$\vert \zeta \vert$ (conversely for small $|v|$). This power-law
decay distribution of the noise is very remarkable and leads to a
dynamics dominated by {\em rare fluctuations}. The occurrence of
large events of amplitude $\vert \zeta \vert \gg 1$ has a small
but non-negligible probability (they take place whenever the
function $v$ gets close to zero). As we will show below, these
rare fluctuations drive the system towards a stationary state that
displays multiscaling properties. Anticipating part of our
numerical results, in Figure 1 we plot the PDF of the chaotic
function $\zeta$ for the Mackey-Glass model, which, as expected
from our theoretical arguments, decays as a power-law
$P_{\zeta}(\zeta) \sim \vert \zeta \vert^{-2.42}$.
\begin{figure}
\centerline{\epsfxsize=7.0cm
\epsfbox{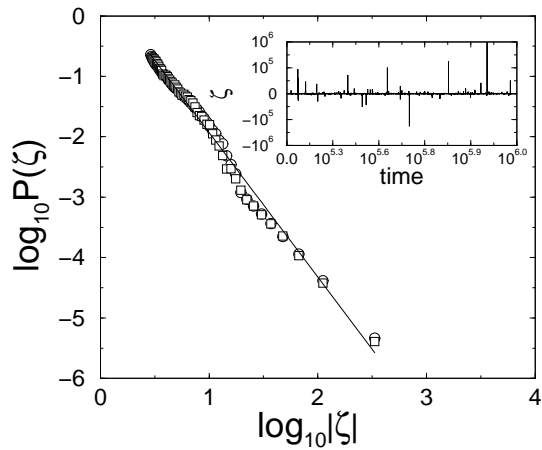}}
\caption{Numerical simulation of the Mackey-Glass model. Inset
shows a typical realization of the chaotic function $\zeta = 1/v$
in time, in which the pronounced non-Gaussian fluctuations are
already visible. Main panel shows the power-law behavior of the
probability density of $\zeta$ obtained for time delays $\tau =
500$ ($\circ$) and $\tau = 2000$ ($\Box$). The line is a
least-mean square fit to the data with slope $-2.42$.}
\end{figure}

The important fact that the surface growth equation that describes
the propagation of disturbances in DDSs, Eq.(\ref{interfeqn}), is
dominated by rare fluctuations strongly suggests that the
universality class of growth to which the surface must belong is
the well-know linear Zhang model:
\begin{equation}
{\partial_{\theta} h} = D\,\nabla^2 h + f_0 + \eta(x,\theta),
\label{zhangmodel}
\end{equation}
which is the representative model of the universality class of
linear surface growth in the presence of power-law noise
\cite{zhang1,krug,baraba,lopez,baraba&stanley}. This model is
expected to describe the dynamics of the Lyapunov vector in DDSs
for large enough systems ({\it i.e.} long enough delay times).
KPZ-like nonlinearities such as $(\nabla h)^2$ are not expected to
appear in the effective surface growth equation, since
Eq.(\ref{interfeqn}) is statistically invariant under the $h \to
-h$ transformation and this symmetry must be preserved. Moreover,
for the class of systems we are interested in this Letter,
$u(x,\theta) = {\rm const}$ and a simple scaling argument shows
that the additive noise in Eq.(\ref{zhangmodel}) must be $\eta
\propto \sqrt{\zeta}$. So, we expect the effective noise term
$\eta$ in Eq.(\ref{zhangmodel}) to be distributed according to a
power-law, $P(\eta) \sim \eta^{-(1+\mu)}$, where the noise index
is $\mu = 2\,(1+\nu)$. Further justification that
Eq.(\ref{interfeqn}) is in the universality class of the linear
Zhang model can be given as follows. By making use of standard
stochastic functional techniques, like Novikov's theorem
\cite{ojalvo&sancho,novikov}, spatial average of the stochastic
drift term $\zeta \, \nabla h$ can be calculated explicitely for
the simpler case of Gaussian noise to obtain $D \nabla^2 h + b
\nabla h $, where the constants $D$ and $b$ are related to
space-time noise correlator integrals (see for instance Ref.\
\cite{ojalvo&sancho}). In the case of noise with zero mean value
and an even spatially homogeneous correlation function $\langle
\eta(x,t) \eta(x',t') \rangle$ one obtains a vanishing drift term
$b=0$.

In the context of surface kinetic roughening,
Eq.(\ref{zhangmodel}) corresponds to the well-known {\em linear}
Zhang model \cite{zhang1,krug,baraba,lopez,baraba&stanley} that
describes the surface height dynamics in the presence of an
uncorrelated power-law distributed noise source. In the following,
we shall briefly review the scaling properties of this growth
model. The height $h(x,t)$ exhibits scale-invariant properties
characterized by the scaling behavior of the surface fluctuations
as measured by the {\em global} width, $W(L,t) = \langle
\overline{[h(x,t) - \overline{h(x,t)}]^2}\rangle^{1/2}$, where the
average is calculated over all $x$ in a system of horizontal size
$L$ (overline) and noise (brackets). Space and time
scale-invariance leads to the scaling behavior of the width,
\begin{equation}
W(L,t)  \sim \left\{
\begin{array}{lcl}
     t^{\beta}     & {\rm if} & t \ll L^z\\
     L^{\alpha} & {\rm if} & t \gg L^z
\end{array}
\right..
\end{equation}
The critical exponents $\alpha$ and $\beta$ are the roughness and
growth exponent, respectively, and characterize the global scaling
behavior of the surface height. The dynamic exponent $z$ is
related with the horizontal extent of the correlation length
$l_{\times} \sim t^{1/z}$ and fulfills the usual relation $z =
\alpha/\beta$. A simple mean-field approximation
\cite{krug,baraba&stanley} provides estimates of the scaling
exponents that depend on the index $\mu$ characterizing the
power-law decay of the noise as follows. The dynamics is diffusive
($z =2$) for any value of $\mu$. However, the global roughness
exponent depends on the noise index, $\alpha = (3-\mu)/\mu$ for
$\mu \leq 2$, when the fluctuations are dominated by rare events,
and $\alpha = 1/2$ for $\mu \geq 2$, where the noise distribution
tail becomes irrelevant and standard Gaussian behavior is
recovered \cite{krug,baraba&stanley}. Another interesting aspect
of roughening dominated by power-law noise is multiscaling
--namely, higher moments of the height-height correlation function
do not scale with the same exponents-- which indicates that the
height distribution is far from Gaussian and dominated by rare
fluctuations of the noise. In particular, multiscaling behavior in
the linear Zhang model occurs for noise indexes in the range $1 <
\mu < 3$ \cite{lopez2}. This behavior is analogous to that
observed in the velocity fluctuations of turbulent fluids
\cite{turbulence} as well as in some solid-on-solid models of
epitaxial growth \cite{krug-mbe,baraba&stanley}. Multiscaling
properties can be proven by calculating the $q$th order
height-height correlation function $G_q(x,t) = \langle
\overline{|h(x_0+x,t) - h(x_0,t)|^q}\rangle^{1/q}$, which scales
as $G_q(x,t) \sim t^{\beta_q}$ for short times and saturates,
$G_q(x) \sim x^{\alpha_q}$, for long enough time
\cite{zhang1,krug,baraba,baraba&stanley}. 
Exponents $\alpha_q$ and $\beta_q$ that depend on the index $q$
are the characteristic fingerprint of multiscaling behavior.

\paragraph{Numerical results.--}
We have carried out extensive numerical
simulations of several chaotic delayed systems in order to compare
with our theoretical findings. We have studied the well-known
class of systems
\begin{equation}
\dot{y}= -a\,y + F (y_\tau), \label{class}
\end{equation}
which have been used in the past for a variety of applications
ranging from biology \cite{mg} to lasers with feedback
\cite{ikeda,roy,goedgebuer}. For the particular choices $F(\rho) =
b\, \rho/(1+\rho^{10})$, $F(\rho) = b\, \sin(\rho-\rho_0)$ and
$F(\rho) = b\, \sin^2(\rho-\rho_0)$, which correspond to the
Mackey-Glass \cite{mg}, Ikeda \cite{ikeda}, and the delayed
feedback model for secure optical communications
\cite{goedgebuer}, respectively. For this class of systems, the
function $u$ is constant, $u=-a$, and the additive noise in
Eq.(\ref{zhangmodel}) is $\eta(x,\theta) \propto \sqrt{1/v}$,
where the nonlinear function $v=dF/d\rho$. In all our numerical
simulations we have used the Adams-Bashforth-Moulton
predictor-corrector scheme \cite{nr}. From the numerical solution
of the corresponding equation for the evolution of the
perturbations of the leading Lyapunov exponent, Eq.(\ref{eqdec1}),
we have constructed the surface
$h(x,\theta)=\log\vert\phi(x,\theta)\vert$ following the procedure
above discussed to get the space-time mapping. After that, the
scaling properties of the resulting surface have been studied. For
the sake of brevity we focus the discussion of numerical results
on the Mackey-Glass model, but similar results were found for the
other two systems studied. The parameters $a=0.1$ and $b=0.2$ are
used in all the results we are presenting here, and simulation
with a time delay varying from ten to a few thousand time units
have been carried out. The region of interest here corresponds to
delays $\tau \gg 17$ for which the Mackey-Glass model is known to
be hyper-chaotic \cite{farmer}.
\begin{figure}
\centerline{\epsfxsize=7.0cm
\epsfbox{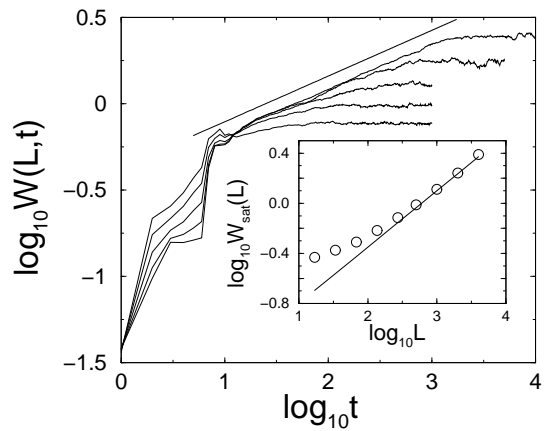}}
\caption{Time evolution of the global surface width in the
Mackey-Glass model for different system sizes $L = 272, 5\times
10^2,10^3,2\times 10^3,4\times 10^3$ (time delays). The straight
line is a guide to the eye and has a slope $\beta = 0.25$. Inset
shows a log-log plot of the stationary $W_{sat}(L)$ {\it vs.} $L$
(flat late regime of the main panel data). The line is a guide to
the eye with slope $\alpha = 0.5$.}
\end{figure}

In Figure 1 (inset) we show a typical run of the simulation for
the Mackey-Glass system with a time delay $\tau = 500$ . The
intermittent behavior of $\zeta$ is already apparent in the form
of large peaks in the inset plot. We recorded the values taken by
$v$ and collected statistics from long runs in order to obtain the
PDF of the noise $\zeta = 1/v$. In Fig. 1 we show our results for
two instances of the time delay $\tau = 500,\, 2000$. We obtain a
power-law distribution $P_{\zeta}(\zeta) \sim |\zeta|^{-2.42}$
that is independent of the time delay used and corresponds to an
exponent $\nu = 0.42 \pm 0.02$. The power spectrum of $\zeta$ (not
shown) is flat indicating that $\zeta$ lacks long-range temporal
correlations. We then expect the noise in the effective surface
growth model (\ref{zhangmodel}) to be power-law distributed with
$\mu = 2\,(1+\nu) = 2.84 \pm 0.04$. This value of the noise index
$\mu > 2$ implies, according to the mean-field prediction for the
linear Zhang model, that the roughness and growth exponent should
be $\alpha = 1/2$ and $\beta = 1/4$, respectively. Very
interestingly, it also implies that one should observe
multiscaling behavior, since $\mu \in (1,3]$.
\begin{figure}
\centerline{\epsfxsize=7.0cm \epsfbox{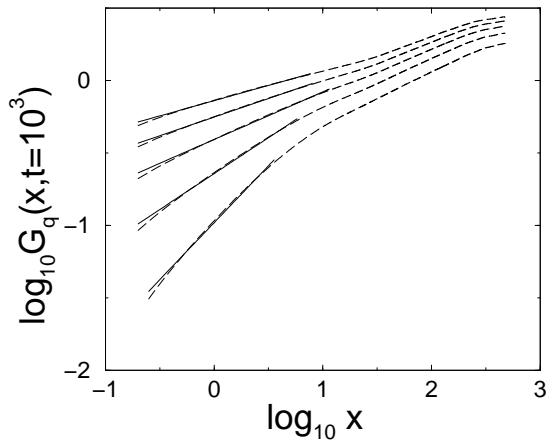}}
\caption{Multiscaling behavior of the stationary $q$th correlation
function of the surface in the Mackey-Glass model (dashed lines)
with a delay $L = 10^3$ for $q=1, 2, 3, 4, 5$ (from bottom to
top). Linear mean least-square fits (solid lines) give an
estimation of the multiscaling roughness exponents as $\alpha_1 =
0.786(6)$, $\alpha_2 = 0.496(2)$, $\alpha_3 = 0.328(1)$, $\alpha_4
= 0.257(1)$, $\alpha_5 = 0.210(1)$.}
\end{figure}

In Figure 2 we plot the time behavior of the surface width for
different system sizes $L$ ({\it i.e.} time delays $\tau$) for the
Mackey-Glass system. From that figure we obtain an estimation of
the growth exponent $\beta = 0.25 \pm 0.03$ in the intermediate
times regime, before a stationary state is reached. The scaling
regime appears after an earlier short transient regime (of about
one decade), and becomes longer for larger system sizes. For the
largest system we used $L=4 \times 10^3$ the scaling regime lasts
for about two decades. After saturation, which occurs at times
$t_{sat} \sim L^z$, the surface height becomes scale-invariant and
the width grows as $W(t \gg t_{sat},L) \sim L^\alpha$, with a
roughness exponent $\alpha = 0.50 \pm 0.05$. From the inset of
Fig. 2, it becomes clear that our estimation of the roughness
exponent fits better in the limit of large system size, showing
that long delays are required to reach the true scaling regime.
These exponent values agree with the theoretical values predicted
for the Zhang model for a noise index $\mu = 2.84
> 2$.

Finally, we have also studied numerically the existence of
multiscaling phenomena. In Figure 3, we plot the $q$th
height-height correlation function for the Mackey-Glass system
with time delay $\tau= 10^3$. Multiscaling behavior is obtained
for any time delay larger than $\tau \approx 17$, which is the
lower bound for chaos to appear in the system. However,
multiscaling becomes clearer in larger systems (longer time
delays), since the multiscaling region has a larger expand. As
predicted by our theoretical arguments, multiscaling behavior is
to be expected whenever $\mu < 3$. Crossover to self-affine
scaling is observed on large scales
as generically expected for surface roughening driven by rare
events (see \cite{baraba&stanley,baraba} for details).

\paragraph{Conclusions.--} Symmetry and scaling considerations
have led us to conclude that the linear Zhang model for surface
roughening driven by power-law noise describes the dynamics of the
Lyapunov vector for long enough time delays. We then argue that
DDSs generically display multiscaling of fluctuations due to the
power-law nature of the driving noise.

The interpretation of DDSs as spatial systems has also been used
in practical applications, in particular for the analysis of
experimental data of $CO_2$ lasers with optical feedback
\cite{arecchi2}. The technique allowed to identify many spacelike
properties in experiments, including defects formation,
transitions between weak and fully developed turbulence, etc.
However, we have established a precise difference between
high-dimensional chaos in delayed systems and in truly extended
systems (like coupled-map lattices, partial differential
equations, etc). Our results also suggest that different
high-dimensional chaotic systems could be divided into a few
universality classes based upon basic symmetries and conservation
laws, akin to what occurs in scale-invariant surface growth.

\begin{acknowledgements}
We thank L. Pesquera for useful comments and discussions. A.D.S.
acknowledges a postdoctoral fellowship from CONICET (Argentina).
Financial support from the MCyT
(Spain) and FEDER under projects BFM2000-0628-C03-02,
BFM2000-1108, BFM2001-0341-C02-02 as well as
project OCCULT IST-2000-29683 from EU are acknowledged.
\end{acknowledgements}

\end{document}